\documentclass{JHEP3}

\title{Zero-size objects in Riemann-Cartan spacetime}
\author{Milovan Vasili\'c, Marko Vojinovi\'c \\
        Institute of Physics, P.O.Box 57, 11001 Belgrade, Serbia \\
        E-mail: \email{mvasilic@phy.bg.ac.yu}, \email{vmarko@phy.bg.ac.yu}}

\abstract{ We use the conservation law of the stress-energy and spin tensors
to study the motion of massive zero-size objects in Riemann-Cartan geometry.
The resultant world line equations turn out to exhibit a novel
spin-curvature coupling. In particular, the spin of the Dirac particle does
not couple to the background curvature. This is a consequence of its truly
zero size which consistently rules out the orbital degrees of freedom. As a
test of consistency, the wave packet solution of the free Dirac equation is
considered. It is shown that the wave packet spin and orbital angular
momentum disappear simultaneously in the zero-size limit. }

\keywords{Classical Theories of Gravity}

\preprint{}

\usepackage{amsmath}

\newcommand{\itGamma}{{\mathit{\Gamma}}}
\newcommand{\del}{\partial}
\newcommand{\cL}{{\cal L}}
\newcommand{\cM}{{\cal M}}
\newcommand{\cO}{{\cal O}}
\newcommand{\cR}{{\cal R}}
\newcommand{\cT}{{\cal T}}
\newcommand{\ds}{\displaystyle}

\newcommand{\lc}{\varepsilon}
\newcommand{\vect}[1]{\vec{#1}}

\newcommand{\orto}{{\scriptscriptstyle\perp}}
\newcommand{\para}{\scriptscriptstyle\parallel}
\newcommand{\Pp}{ {P_{\para}}\vphantom{P} }
\newcommand{\Pn}{ {P_{\orto}}\vphantom{P} }
\newcommand{\cfl}[2]{{\textstyle {{#1}\brace {#2}}}}
\newcommand{\h}{h}

\begin{document}

\section{\label{IntroductionSection}Introduction}

The problem of particle motion in backgrounds of nontrivial geometry is
usually addressed by using some form of the Mathisson-Papapetrou method
\cite{Mathisson1937, Papapetrou1951}. One starts with the covariant
conservation law of the stress-energy and spin tensors of matter fields, and
analyzes it under the assumption that matter is highly localized. In the
lowest, single-pole approximation, the moving matter is viewed as a point
particle. In the pole-dipole approximation, its non-zero size is taken into
account.

The results found in literature can be summarized as follows. Spinless
particles in the single-pole approximation obey the geodesic equation. In
the pole-dipole approximation, the rotational angular momentum of the
localized matter couples to spacetime curvature, and produces geodesic
deviations \cite{Mathisson1937, Papapetrou1951, Tulczyjew1959, Taub1964,
Dixon}. If the particles have spin, the curvature couples to
the total angular momentum, and the torsion to the spin alone
\cite{Trautman1972, Hehl1976a, Yasskin1980, Nomura1991, Nomura1992}.

What we are interested in is a consistent single-pole analysis of spinning
particles in spacetimes with curvature and torsion. This is motivated by the
observation that single-pole approximation eliminates the influence of
particle thickness, and allows the derivation of the pure spin-curvature
coupling. In fact, this is the only way to see the influence of curvature on
the spin part of the total angular momentum. The ambiguous algebraic
decomposition of the total angular momentum into spin and orbital
contributions are of no help. What we need is a truly zero-size object. As
it turns out, the existing literature on the subject  does not have this
sort of prediction.

The results that we have obtained are summarized as follows. Trajectories of
spinning zero-size massive particles generally deviate from the geodesic
lines. The deviation is due to the spin-curvature and spin-torsion
couplings. These turn out to be different from what has been believed so
far. In particular, the spin of the Dirac point particle does not couple to
the curvature. If it is viewed as a wave packet solution of the Dirac
equation, it does not couple to the torsion either. In fact, the wave packet
spin and orbital angular momentum disappear simultaneously in the zero-size
limit. We can say that Dirac point particles behave as spinless objects.

The layout of the paper is as follows. In section \ref{SinglePoleSection},
we define the conservation law of the stress-energy and spin tensors, and
introduce the necessary geometric notions. The algebraic part of the
conservation equations is solved in terms of the independent variables---the
spin tensor and the symmetric part of the stress-energy tensor. After the
brief recapitulation of the covariant multipole formalism, we define the
single-pole approximation for the independent variables, only. Section
\ref{EOMSection} is devoted to the derivation of the particle world line
equations. The actual derivation is only sketched, as the method has already
been analyzed in detail in \cite{Vasilic2007}. The resulting equations of
motion are compared to the pole-dipole equations found in literature
\cite{Yasskin1980, Nomura1991}. As it turns out, they coincide up to a
constraint that fixes the form of the spin tensor. This constraint is a
consequence of our single-pole approximation, and has striking consequences
on the dynamics of the Dirac particle. In section \ref{DiracSection}, we
discuss the important case of totally antisymmetric spin tensor, and obtain
a surprising result that spin of the Dirac particle does not couple to the
background curvature. To check the consistency of our single-pole
approximation, the wave packet solution of the free Dirac equation is
analyzed. It is demonstrated that the wave packet spin and orbital angular
momentum disappear simultaneously in the zero-size limit. In section
\ref{ConclusionSection}, we give our final remarks.

Conventions in this paper are the following. Greek indices from the middle
of the alphabet, $\mu,\nu,\dots$, are the spacetime indices, and run over
$0,1,2,3$. The indices from the beginning of the Greek alphabet,
$\alpha,\beta,\dots$, take values $1,2,3$. The spacetime coordinates
are denoted by $x^{\mu}$, the generic metric is denoted by $g_{\mu\nu}(x)$,
and $\eta_{\mu\nu}$ stands for the Minkowski metric. The signature
convention is $(-+++)$.

\section{\label{SinglePoleSection}The single-pole approximation}

We begin with the covariant conservation of the fundamental matter currents
--- stress-energy tensor $\tau^{\mu}{}_{\nu}$, and spin tensor
$\sigma^{\lambda}{}_{\mu\nu}$:
\begin{subequations} \label{ZakoniOdrzanja}
\begin{equation} \label{ZakonOdrzanjaTau}
\left( D_{\nu} + \cT^{\lambda}{}_{\nu\lambda} \right) \tau^{\nu}{}_{\mu} =
\tau^{\nu}{}_{\rho} \cT^{\rho}{}_{\mu\nu} + \frac{1}{2}
\sigma^{\nu\rho\sigma}\cR_{\rho\sigma\mu\nu},
\end{equation}
\begin{equation} \label{ZakonOdrzanjaSigma}
\left( D_{\nu} + \cT^{\lambda}{}_{\nu\lambda}
\right)\sigma^{\nu}{}_{\rho\sigma} = \tau_{\rho\sigma} - \tau_{\sigma\rho}.
\end{equation}
\end{subequations}
Here, $D_{\nu}$ is the covariant derivative with the nonsymmetric connection
$\itGamma^{\lambda}{}_{\mu\nu}$, which acts on a vector $v^{\mu}$ according
to the rule $D_{\nu} v^{\mu} \equiv \del_{\nu}v^{\mu} +
\itGamma^{\mu}{}_{\lambda\nu}v^{\lambda}$. The torsion
$\cT^{\lambda}{}_{\mu\nu}$, and curvature $\cR^{\mu}{}_{\nu\rho\sigma}$ are
defined in the standard way:
$$
\cT^{\lambda}{}_{\mu\nu} \equiv \itGamma^{\lambda}{}_{\nu\mu} -
\itGamma^{\lambda}{}_{\mu\nu}, \qquad \cR^{\mu}{}_{\nu\rho\sigma} \equiv
\del_{\rho} \itGamma^{\mu}{}_{\nu\sigma} - \del_{\sigma}
\itGamma^{\mu}{}_{\nu\rho} + \itGamma^{\mu}{}_{\lambda\rho}
\itGamma^{\lambda}{}_{\nu\sigma} - \itGamma^{\mu}{}_{\lambda\sigma}
\itGamma^{\lambda}{}_{\nu\rho}.
$$
The derivative $D_{\lambda}$ is assumed to satisfy the metricity condition,
$D_{\lambda}g_{\mu\nu}=0$. As a consequence, the connection
$\itGamma^{\lambda}{}_{\mu\nu}$ is split into the Levi-Civita connection
$\cfl{\lambda}{\mu\nu}$, and the contorsion $K^{\lambda}{}_{\mu\nu}$:
$$
\itGamma^{\lambda}{}_{\mu\nu} = \cfl{\lambda}{\mu\nu} +
K^{\lambda}{}_{\mu\nu}, \qquad K^{\lambda}{}_{\mu\nu} \equiv -\frac{1}{2}
\left( \cT^{\lambda}{}_{\mu\nu} - \cT_{\nu}{}^{\lambda}{}_{\mu} +
\cT_{\mu\nu}{}^{\lambda} \right).
$$
We shall also introduce the Riemannian covariant derivative $\nabla_{\mu}
\equiv D_{\mu}(\itGamma\to \{ \} )$, and the Riemannian curvature tensor
$R^{\mu}{}_{\nu\rho\sigma} \equiv \cR^{\mu}{}_{\nu\rho\sigma}(\itGamma \to
\{ \} )$. The relation connecting the two curvature tensors reads:
$$
\cR^{\mu}{}_{\nu\lambda\rho} = R^{\mu}{}_{\nu\lambda\rho} + 2
\nabla_{[\lambda} K^{\mu}{}_{\nu\rho]} + 2 K^{\mu}{}_{\sigma[\lambda}
K^{\sigma}{}_{\nu\rho]},
$$
where the indices in square brackets are antisymmetrized.

Given the system of conservation equations (\ref{ZakoniOdrzanja}), one finds that
the second one has no dynamical content. Indeed, the antisymmetric part of
stress-energy tensor is completely determined by the spin tensor. One can use
(\ref{ZakonOdrzanjaSigma}) to eliminate $\tau^{[\mu\nu]}$ from the equation
(\ref{ZakonOdrzanjaTau}), and thus obtain the conservation equation, in which
only $\tau^{(\mu\nu)}$ and $\sigma^{\lambda\mu\nu}$ components appear. The
resulting equation reads:
\begin{equation} \label{GlavniZakonOdrzanja}
\nabla_{\nu}\left( \tau^{(\mu\nu)} + \frac{1}{2} K_{\lambda\rho}{}^{\mu}
\sigma^{\nu\lambda\rho} - K^{[\mu}{}_{\lambda\rho} \sigma^{\rho\lambda\nu]}
- \nabla_{\rho} \sigma^{(\mu\nu)\rho} \right) = \frac{1}{2}
\sigma_{\nu\rho\lambda}\nabla^{\mu} K^{\rho\lambda\nu}.
\end{equation}
This will be the starting point of the derivation of the particle world line
equations.

Let us now introduce the multipole formalism, which is necessary for the
derivation. It has been shown in Refs. \cite{Vasilic2006, Vasilic2007} that an
exponentially decreasing function can be expanded into a series of
$\delta$-function derivatives. For example, a scalar $V(x)$, well localized
around the line $\cM$, can be written in a manifestly covariant way as
\begin{equation} \label{DeltaSeries}
V(x)=\int_{\cM}\! d\tau \!\left[ M(\tau)\frac{\delta^{(4)}(x-z)}{\sqrt{-g}} -
\nabla_{\rho}\left(M^{\rho}(\tau)\frac{\delta^{(4)}(x-z)}{\sqrt{-g}}\right) +
\cdots \right].
\end{equation}
Here, $\cM$ is a timelike line $x^{\mu} = z^{\mu}(\tau)$ parametrized by the
proper distance, $d\tau^2=g_{\mu\nu}dz^{\mu}dz^{\nu}$, and the coefficients
$M(\tau)$, $M^{\rho}(\tau)$, ... are spacetime tensors called multipole
coefficients. It has been shown in Ref. \cite{Vasilic2007} that one may
truncate the series in a covariant way in order to approximate the
description of matter. Truncation after the leading term is called
\emph{single-pole} approximation, truncation after the second term is called
\emph{pole-dipole} approximation. The physical interpretation of these
approximations is the following. In the single-pole approximation, one
assumes that the particle has no thickness, which means that matter is
localized in a point. All higher approximations, including pole-dipole,
allow for the nonzero thickness, and thus, for the nontrivial internal motion.

Apart from being covariant with respect to diffeomorphisms, the series
(\ref{DeltaSeries}) possesses two extra gauge symmetries. The first is a
consequence of the fact that that there are redundant coefficients in this
decomposition. Indeed, only three out of four $\delta$-functions in each
term of the multipole expansion (\ref{DeltaSeries}) are effective in
modeling particle trajectory in $4$-dimensional spacetime. The extra
$\delta$-function and the extra integration are introduced only to
covariantize the expressions. The derivatives parallel to the world line are
integrated out, as they should, considering the fact that matter is not
localized in time. As a consequence, the parallel components of the
multipole coefficients $M^{\rho}$, $M^{\rho\lambda}$, ... effectively
disappear. It has been shown in Ref. \cite{Vasilic2007} that the
corresponding gauge symmetry, named \emph{extra symmetry}~1, in the
pole-dipole approximation reads:
$$
\delta_1 M = \nabla \epsilon \,, \qquad
\delta_1 M^{\rho} = u^{\rho}\epsilon \,.
$$
Here, $u^{\mu}\equiv dz^{\mu}/d\tau$ is the particle $4$-velocity, $\epsilon
(\tau)$ is a gauge parameter, and $\nabla$ stands for the Riemannian covariant
derivative along the particle trajectory (\,$\nabla v^{\mu}=dv^{\mu}/d\tau
+\cfl{\mu}{\lambda\rho}v^{\lambda}u^{\rho}$\,). We see that the parallel
component of $M^{\rho}$ transforms as $\delta_1
\left(M^{\rho}u_{\rho}\right)=-\epsilon\,$, and can be gauged away. In fact, one
can show that the parallel components of the higher multipoles are also pure
gauge. In the gauge fixed multipole expansion, the only derivatives that appear
are those orthogonal to the world line.

The second extra symmetry stems from the fact that the choice of the line
$x^{\mu}=z^{\mu}(\tau)$ in the expansion (\ref{DeltaSeries}) is arbitrary.
If we use another line, let us say $x^{\mu} = z'^{\mu}(\tau)$, the
coefficients $M$, $M^{\rho}$, ... will change to $M'$, $M'^{\rho}$, ...
while leaving the scalar function $V(x)$ invariant. The transformation law
of the $M$-coefficients, generated by the replacement $z^{\mu}\to z'^{\mu}$,
defines the gauge symmetry that we call \emph{extra symmetry}~2.

The extra symmetry 2 is an exact symmetry of the full expansion
(\ref{DeltaSeries}), but only approximate symmetry of the truncated series.
In the pole-dipole approximation, it has the form
$$
\delta_2 z^{\mu} = \epsilon^{\mu}\,,\qquad
\delta_2 M = M u_{\rho}\nabla\epsilon^{\rho}\,,\qquad
\delta_2 M^{\rho} = -M \epsilon^{\rho}\,,
$$
provided the $M$-coefficients are subject to the hierarchy
$$
M = \cO_0\,, \quad
M^{\rho} = \cO_1\,, \quad
M^{\rho\lambda} = \cO_2 \,,\ \dots \,,
$$
and the free parameters $\epsilon^{\mu}(\tau)$ satisfy $\epsilon^{\mu} =
\cO_1$. Here, $\cO_n$ stands for the order of smallness, and the condition
$\epsilon^{\mu} = \cO_1$ ensures that the order of truncation is not
violated by the action of the symmetry transformations \cite{Vasilic2007}.
In the pole-dipole and higher approximations, fixing the gauge of extra
symmetry 2 defines the particle centre of mass. In the single-pole
approximation, the extra symmetry 2 is trivial.

Now, we shall replace the general function $V(x)$ with the stress-energy and
spin tensors of the localized matter. In order to describe a strict point
particle, we choose $\tau^{(\mu\nu)}$ and $\sigma^{\lambda\mu\nu}$ in the
form
\begin{subequations} \label{SinglePoleAproksimacija}
\begin{equation} \label{DeltaRazvojZaTau}
\tau^{(\mu\nu)}  = \int_{\cM} d\tau\, b^{\mu\nu}(\tau)
\frac{\delta^{(D)}(x-z)}{\sqrt{-g}},
\end{equation}
\begin{equation} \label{DeltaRazvojZaSigma}
\sigma^{\lambda\mu\nu} = \int_{\cM} d\tau\, c^{\lambda\mu\nu}(\tau)
\frac{\delta^{(D)}(x-z)}{\sqrt{-g}}\,,
\end{equation}
\end{subequations}
where $b^{\mu\nu}(\tau)$ and $c^{\lambda\mu\nu}(\tau)$ are the corresponding
multipole coefficients. We emphasize here that this is not how single-pole
approximation is defined in the existing literature \cite{Yasskin1980,
Nomura1991}. There, the antisymmetric part of stress-energy tensor
$\tau^{\mu\nu}$ has also been treated in the single-pole manner. As
$\tau^{[\mu\nu]}$ is not an independent variable, this imposed unnecessary
constraints on $\sigma^{\lambda\mu\nu}$. In particular, the spin of the
Dirac particle was ruled out. To overcome this problem, the authors of Ref.
\cite{Nomura1991} abandoned single-pole in favour of pole-dipole
approximation. Their subsequent limit of vanishing orbital angular momentum
should have brought them back to the single-pole regime. In what follows,
however, we shall demonstrate that it is not quite so, and that such a limit
is not equivalent to the single-pole approximation as defined in
(\ref{SinglePoleAproksimacija}).

\section{\label{EOMSection}Equations of motion}

The particle equations of motion are derived in the following way. We insert
(\ref{SinglePoleAproksimacija}) into (\ref{GlavniZakonOdrzanja}), and solve
for the unknown variables $z(\tau)$, $b^{\mu\nu}(\tau)$ and
$c^{\lambda\mu\nu}(\tau)$. The algorithm for solving this type of equation
is discussed in detail in \cite{Vasilic2006, Vasilic2007}, and here we only
sketch it. The first step is to multiply the equation
(\ref{GlavniZakonOdrzanja}) with an arbitrary spacetime function
$f_{\mu}(x)$, and integrate over the spacetime. The resulting equation
depends on the function $f_{\mu}$ and its first and second covariant
derivatives, evaluated on the line $x^{\mu}=z^{\mu}(\tau)$:
$$
\int d\tau \Big[c^{(\mu\nu)\rho}f_{\mu ;\nu\rho} +
\Big(b^{\mu\nu} - K^{[\mu}{}_{\lambda\rho} c^{\rho\lambda\nu]} +
\frac{1}{2}K_{\lambda\rho}{}^{\mu}c^{\nu\lambda\rho}\Big)f_{\mu ;\nu} +
\frac{1}{2}c_{\nu\rho\lambda}\Big(\nabla^{\mu}K^{\rho\lambda\nu}\Big)
f_{\mu}\Big]=0\,,
$$
where $f_{\mu;\nu}\equiv (\nabla_{\nu}f_{\mu})_{x=z}$, $f_{\mu;\nu\rho} \equiv
(\nabla_{\rho}\nabla_{\nu} f_{\mu})_{x=z}$. Owing to the arbitrariness of the
function $f_{\mu}(x)$, the terms proportional to its independent derivatives
separately vanish. To find the independent derivatives of the test function
$f_{\mu}$, we make use of the particle $4$-velocity $u^{\mu}\equiv
dz^{\mu}/d\tau$, and the Riemannian covariant derivative along the particle
trajectory $\nabla$. The $4$-velocity $u^{\mu}$ is normalized as
$u^{\mu}u_{\mu}=-1$, and the action of $\nabla$ on a vector field $v^{\mu}(\tau)$
is defined by $\nabla v^{\mu}\equiv dv^{\mu}/d\tau
+\cfl{\mu}{\lambda\rho}v^{\lambda}u^{\rho}$. Next, we decompose the derivatives
of the vector field $f_{\mu}(x)$ into components orthogonal and parallel to the
world line $x^{\mu}=z^{\mu}(\tau)$:
\begin{subequations} \label{jna21}
\begin{equation} \label{jna21a}
f_{\mu;\lambda} = f^{\orto}_{\mu\lambda} - u_{\lambda} \nabla f_{\mu} \, ,
\end{equation}
\begin{equation} \label{jna21b}
f_{\mu;(\lambda\rho)} = f^{\orto}_{\mu\lambda\rho} - 2
\h^{\orto}_{\mu(\lambda} u_{\rho)} + \h_{\mu } u_{\lambda} u_{\rho} \, ,
\end{equation}
\begin{equation} \label{jna21c}
f_{\mu;[\lambda\rho]} = \frac{1}{2} {R^{\sigma}}_{\mu\lambda\rho}
f_{\sigma}\, .
\end{equation}
\end{subequations}
Here, the orthogonal and parallel components are obtained by using the projectors
\begin{equation} \label{jna22}
\Pn^{\mu}_{\nu} = \delta^{\mu}_{\nu} + u^{\mu}u_{\nu}, \qquad
\Pp^{\mu}_{\nu} = - u^{\mu} u_{\nu} .
\end{equation}
More precisely, $f^{\orto}_{\mu\lambda} = \Pn^{\sigma}_{\lambda}
f_{\mu;\sigma}$, $f^{\orto}_{\mu\lambda\rho} = \Pn^{\sigma}_{\lambda}
\Pn^{\nu}_{\rho} f_{\mu;(\sigma\nu)}$, $\h^{\orto}_{\mu\lambda} =
\Pn^{\sigma}_{\lambda} u^{\nu} f_{\mu;(\sigma\nu)}$ and $\h_{\mu} =
u^{\sigma} u^{\nu} f_{\mu;(\sigma\nu)}$. Direct calculation yields
\begin{equation} \label{jna23}
\begin{array}{lcl}
\h_{\mu} & = & \nabla \nabla f_{\mu} - (\nabla u^{\nu}) f^{\orto}_{\mu\nu}
\, , \\ \h^{\orto}_{\mu\rho} & = & \ds \Pn^{\nu}_{\rho} \nabla
f^{\orto}_{\mu\nu} - (\nabla u_{\rho}) \nabla f_{\mu} + \frac{1}{2}
\Pn^{\lambda}_{\rho} u^{\nu} {R^{\sigma}}_{\mu\nu\lambda} f_{\sigma} \, , \\
\end{array}
\end{equation}
which tells us that the only independent derivatives on the line
$x^{\mu}=z^{\mu}(\tau)$ are $f_{\mu}$, $f^{\orto}_{\mu\nu}$ and
$f^{\orto}_{\mu\nu\rho}$. We can now use (\ref{jna21}) and (\ref{jna23}) to
group the coefficients in terms proportional to the independent
derivatives of $f_{\mu}$. The obtained equation has the following general
structure:
$$
\int_{\cM}d\tau \Big[ X^{\mu\nu\rho}f^{\perp}_{\mu\nu\rho} +
X^{\mu\nu}f^{\perp}_{\mu\nu} + X^{\mu}f_{\mu} \Big] =0,
$$
where $X^{\mu\nu\rho}$, $X^{\mu\nu}$ and $X^{\mu}$ are composed of various
combinations of multipole coefficients $b^{\mu\nu}$ and $c^{\lambda\mu\nu}$,
external fields $K^{\lambda}{}_{\mu\nu}$ and $R^{\mu}{}_{\nu\rho\sigma}$, and
their derivatives. In all the expressions, the external fields are evaluated on
the world line $x^{\mu}=z^{\mu}(\tau)$. Owing to the fact that $f_{\mu}$,
$f^{\orto}_{\mu\nu}$ and $f^{\orto}_{\mu\nu\rho}$ are independent functions on
the world line, we deduce that the $X$-terms must separately vanish.
The equation $X^{\mu\nu\rho}=0$ has a simple algebraic form
$$
\Pn^{\mu}_{\lambda} \Pn^{\nu}_{\sigma} c^{(\lambda\sigma)\rho} = 0\,,
$$
which is easily solved for $c^{\lambda\mu\nu}$. This yields
$$
c^{\lambda\mu\nu} = 2 u^{\lambda}s^{\mu\nu} + s^{\lambda\mu\nu}\,,
$$
where $s^{\mu\nu} \equiv - s^{\nu\mu}$ and $s^{\lambda\mu\nu} \equiv -
s^{\lambda\nu\mu} \equiv s^{\nu\lambda\mu}$ are totally antisymmetric, but
othervise free parameters. The equations $X^{\mu\nu}=0$ and $X^{\mu}=0$ are much
more complicated. The procedure goes as follows. First, we use the above
decomposition of $c^{\lambda\mu\nu}$ to perform a similar split of the
$b^{\mu\nu}$ coefficients. A new free parameter $m(\tau)$ appears to characterize
the leading term of $b^{\mu\nu}$. Then, the equations $X^{\mu\nu}=0$ and
$X^{\mu}=0$ are rewritten in terms of the undetermined parameters $m$,
$s^{\mu\nu}$ and $s^{\lambda\mu\nu}$, and properly rearranged. Skipping the
details of the diagonalization procedure, which has thoroughly been demonstrated
in Ref. \cite{Vasilic2007}, we display the final result:
\begin{itemize}
\item the world line equation
\begin{subequations} \label{Rezultati}
\begin{equation} \label{PrvaJnaKretanja}
\nabla \Big[ mu^{\mu} + 2u_{\rho} \left( \nabla s^{\mu\rho} + D^{\mu\rho}
\right) \Big] - u^{\nu} s^{\rho\sigma} R^{\mu}{}_{\nu\rho\sigma} =
\frac{1}{2}c_{\nu\rho\lambda} \nabla^{\mu}K^{\rho\lambda\nu},
\end{equation}
\item the spin precession equation
\begin{equation} \label{DrugaJnaKretanja}
\Pn^{\mu}_{\rho} \Pn^{\nu}_{\sigma}\left( \nabla s^{\rho\sigma} +
D^{\rho\sigma} \right) =0,
\end{equation}
\item the stress-energy coefficients
\begin{equation} \label{KomponenteTEI}
b^{\mu\nu} = m u^{\mu}u^{\nu} + 2u_{\lambda}u^{(\mu}\nabla s^{\nu)\lambda} -
\frac{1}{2}K_{\lambda\rho}{}^{(\mu} c^{\nu)\lambda\rho},
\end{equation}
\item the spin tensor coefficients
\begin{equation} \label{SveKomponenteC}
c^{\lambda\mu\nu} = 2 u^{\lambda}s^{\mu\nu} + s^{\lambda\mu\nu}.
\end{equation}
\end{subequations}
\end{itemize}
In these equations, the scalar $m(\tau)$, and the totally antisymmetric
tensors $s^{\mu\nu}(\tau)$ and $s^{\lambda\mu\nu}(\tau)$ are free
parameters. They determine the stress-energy and spin tensors via
(\ref{KomponenteTEI}) and (\ref{SveKomponenteC}).
The shorthand notation
$$
D^{\mu\nu} \equiv K^{[\mu}{}_{\lambda\rho} c^{\rho\lambda\nu]} +
\frac{1}{2}K_{\lambda\rho}{}^{[\mu} c^{\nu]\rho\lambda}
$$
is introduced to simplify the cumbersome expressions.

The obtained single-pole equations differ from the known pole-dipole result
\cite{Trautman1972, Hehl1976a, Yasskin1980, Nomura1991, Nomura1992} by the
presence of the constraint (\ref{SveKomponenteC}). It is a consequence of
our assumption that the particle has no thickness, and therefore no orbital
degrees of freedom. In the existing literature, an analogous but more
restrictive constraint appears in this regime \cite{Hehl1976a, Nomura1991}.
This is because the antisymmetric part of the stress-energy tensor
$\tau^{[\mu\nu]}$ has been treated as an independent variable, in spite of
the restriction (\ref{ZakonOdrzanjaSigma}). In our approach, the only
independent variables are $\sigma^{\lambda\mu\nu}$ and $\tau^{(\mu\nu)}$,
and the resulting constraint (\ref{SveKomponenteC}) is not so strong. In
particular, it does not rule out the free Dirac field, or any other massive
elementary field. Indeed, the formula $c^{\lambda\mu\nu} =
u^{\lambda}s^{\mu\nu} + \frac{1}{s}u^{[\mu}s^{\nu]\lambda}$ for the spin
tensor of the elementary particle of spin $s$ (see Ref. \cite{Nomura1992})
is a special case of (\ref{SveKomponenteC}).

In what follows, we shall examine the special case of spin $1/2$ pointlike
matter. Surprisingly, we shall discover that spin $1/2$ does not couple to
the curvature, leading to geodesic trajectories in torsionless spacetimes.

\section{\label{DiracSection}The Dirac particle}

The basic property of Dirac matter is the total antisymmetry of its spin
tensor $\sigma^{\lambda\mu\nu}$. As a consequence, the coefficients
$c^{\lambda\mu\nu}$ are also totally antisymmetric, and the constraint
(\ref{SveKomponenteC}) implies
\begin{equation}
s^{\mu\nu}=0.
\end{equation}
The vanishing of the $s^{\mu\nu}$ component of the spin tensor has far
reaching consequences. First, we see that the spin-curvature and spin-orbit
couplings disappear from the world line equation (\ref{PrvaJnaKretanja}).
Second, the spin precession equation (\ref{DrugaJnaKretanja}) becomes a
constraint equation. If we define the spin vector $s^{\mu}$ by
$s^{\mu\nu\rho} \equiv e^{\mu\nu\rho\lambda}s_{\lambda}$, and the axial
component of the contorsion $K^{\mu}$ as $K^{\mu} \equiv
e^{\mu\nu\rho\lambda} K_{\nu\rho\lambda}$, where $e^{\mu\nu\rho\sigma}$ is
the covariant totally antisymmetric Levi-Civita tensor, the equations
(\ref{Rezultati}) become
\begin{subequations} \label{jna6}
\begin{equation} \label{jna6a}
\nabla\left( mu^{\mu} + K^{[\mu}s^{\nu]} u_{\nu} \right) + \frac{1}{2}
s^{\nu} \nabla^{\mu}K_{\nu} =0 ,
\end{equation}
\begin{equation} \label{jna6b}
K_{\perp}^{[\mu}s_{\perp}^{\nu]} = 0.
\end{equation}
\end{subequations}
As we can see, the spin couples only to the axial component of the
contorsion, which means that \emph{Dirac point particles follow geodesic
trajectories in torsionless spacetimes}. At the same time, the absence of
torsion trivializes the equation (\ref{jna6b}), and no information on the
behavior of the spin vector is available. If the background torsion has
nontrivial axial component, a geodesic deviation appears, but also a very
strong constraint on the spin vector. Indeed, the equation (\ref{jna6b})
implies that the orthogonal component of $s^{\mu}$ always orients itself
along the background direction $K_{\perp}^{\mu}$. This unusual behavior
suggests that the spin vector of the Dirac point particle might be zero,
after all. In fact, the world line equations (\ref{Rezultati}) are derived
under very general assumptions of the existence of pointlike solutions in an
arbitrary field theory. They do not care about peculiarities of specific
theories or specific types of localized solutions. In what follows, we shall
analyze the wave packet solutions of the flat space Dirac equation, with the
idea to check if they can be viewed as point particles.

Let us construct an example. We start with the free Dirac Lagrangian
$$
\cL = \frac{i}{2} \left[ \bar{\psi}\gamma^{\mu}\del_{\mu} \psi - \left(
\del_{\mu} \bar{\psi} \gamma^{\mu} \psi \right) \right] - m\bar{\psi}\psi\,,
$$
where Dirac $\gamma$-matrices satisfy the usual anticommutation relations
$\{ \gamma^{\mu}, \gamma^{\nu} \} = -2\eta^{\mu\nu}$, and are used in their
conventional representation ($\gamma_5 \equiv i
\gamma^0\gamma^1\gamma^2\gamma^3$). Then, we construct a wave packet. The
wave packet is a solution which is well localized in space, but resembles a
plane wave inside. To be viewed as a particle, its size $\ell$ is considered
in the limit $\ell\to 0$. At the same time, the particle stability is
achieved in the limit $\lambda / \ell \to 0$, where $\lambda$ is its
wavelength. We construct it as follows. At the initial moment $x^0=0$, we
choose the configuration
$$
\psi(\vect{r},0) \equiv Ae^{-\frac{r^2}{\ell^2}}\,\psi_p(\vect{r},0)\,,
$$
where
$$
\psi_p(x) \equiv \sqrt{\frac{k^0+m}{2m}}
\left[
\begin{array}{c}
1 \\ 0 \\ \frac{k^3}{k^0+m} \\ 0 \\
\end{array}
\right]
e^{ik_{\mu}x^{\mu}}
$$
is the plane-wave solution of the Dirac equation $\left(
i\gamma^{\mu}\del_{\mu} -m \right) \psi =0$. It propagates along the
$x^3$-axis ($k^1=k^2=0$, $k^0 \equiv \sqrt{m^2 + (k^3)^2}$), and is
polarized upwards, for convenience. The exponential function $\exp
(-r^2/\ell^2)$ cuts out a small piece of the plane wave, and defines its
size $\ell$, while $A$ is the overall amplitude of the packet. The
wavelength $\lambda$ is proportional to $1/|\vect{k}|$. Using the Dirac
equation, we can calculate time derivatives and thereby determine time
evolution of this packet. In fact, we only need first time derivatives, as
neither $\tau^{\mu\nu}$ nor $\sigma^{\lambda\mu\nu}$ depend on higher
derivatives:
\begin{equation} \label{jna7}
\tau_{\mu\nu} = i \left[ \bar{\psi} \gamma_{\mu} \del_{\nu}\psi - \left(
\del_{\nu} \bar{\psi} \right) \gamma_{\mu}\psi \right]
-2\eta_{\mu\nu}\cL\,,\qquad \sigma^{\lambda\mu\nu} = \lc^{\lambda\mu\nu\rho}
\bar{\psi} \gamma_5 \gamma_{\rho} \psi\,.
\end{equation}
The wave packet expressions of these currents at $x^0=0$ are obtained
straightforwardly:
\begin{subequations} \label{RnjeTalasnogPaketa}
\begin{equation}\label{TauNulaNulaItauTriTri}
\tau^{(00)} = -2 |A|^2 e^{-\frac{2r^2}{\ell^2}} \frac{(k^0)^2}{m} , \qquad
\tau^{(33)} = -2 |A|^2 e^{-\frac{2r^2}{\ell^2}} \frac{(k^3)^2}{m} ,
\end{equation}
\begin{equation}\label{TauNulaAlfa}
\tau^{(0\alpha)} = -2 |A|^2 e^{-\frac{2r^2}{\ell^2}} \frac{k^0}{m} \left(
k^3\eta^{3\alpha} - \frac{x_{\beta}}{\ell^2} \lc^{\alpha\beta 3} \right) ,
\end{equation}\label{Sigma123}
\begin{equation}
\sigma^{123} = - |A|^2 e^{-\frac{2r^2}{\ell^2}} \frac{k^3}{m}, \qquad
\sigma^{012} = - |A|^2 e^{-\frac{2r^2}{\ell^2}} \frac{k^0}{m},
\end{equation}
\end{subequations}
where only non-vanishing components are displayed. Now, we want to rewrite
the currents (\ref{RnjeTalasnogPaketa}) as a series of $\delta$-function
derivatives. We first fix diffeomorphisms by imposing the condition
$g_{\mu\nu}=\eta_{\mu\nu}$, and extra symmetry 1 by keeping only spatial
components of the M coefficients ($M^0=M^{0\rho}=\cdots =0$). The
decomposition formula (\ref{DeltaSeries}) is thereby reduced to
$$
V(x)=\int_{\cM}\! d\tau \!\left[ M \delta^{(4)}(x-z) -
\partial_{\alpha}\left(M^{\alpha}\delta^{(4)}(x-z)\right) +
\cdots \right].
$$
In general, the line $x^{\mu}=z^{\mu}(\tau)$ is arbitrary, but the simplest
expressions are obtained if it coincides with the wave packet trajectory.
Thus, we choose $z^1=z^2=0$ in accordance with the the fact that the packet
propagates along the $x^3$ axis. As for the $z^3$ component, we do not need
its full $\tau$ dependence because we are only interested in the packet
behaviour at $x^0=0$. There, the proper length $\tau$ is chosen in
accordance with $z^0(0)=z^3(0)=0$, which is sufficient for the proper
definition of the $\delta$-expansion at $x^0=0$.

The multipole coefficients are obtained by multiplying $V(x)$ by a number of
$(x^{\alpha}-z^{\alpha})$ factors, and integrating over the $3$-space. In our
example, the simple integration of (\ref{RnjeTalasnogPaketa}) yields the monopole
coefficients $b^{\mu\nu}$ and $c^{\lambda\mu\nu}$, while multiplication with
$(x^{\alpha}-z^{\alpha})$, and subsequent integration gives the dipole
coefficients $b^{\mu\nu\alpha}$. The resultant non-zero monopoles are
$$
b^{00} = a\, \ell^3 (k^0)^2 \,, \qquad
b^{33} = a\, \ell^3 (k^3)^2 \,, \qquad
b^{03} = a\, \ell^3 k^0 k^3 \,,
$$
$$
c^{123} = \frac{a}{2}\, \ell^3 k^3\,, \qquad
c^{012} = \frac{a}{2}\, \ell^3 k^0\,,
$$
while there are only two non-zero dipoles,
$$
b^{012} = - b^{021} = \frac{a}{4}\, \ell^3 k^0 \,.
$$
The higher multipoles are of the order $\ell^5(k^3)^2$ or higher. Here, $a(k)$ is
the overall factor whose explicit form is not needed in the subsequent
discussion.

Now, we shall consider the single-pole limit $\ell\to 0$, while respecting the
wave packet stability condition $\lambda\ll\ell$. First, we choose the overall
amplitude $a(k)$ in the form
$$
a(k) \sim \frac{\lambda^2}{\ell^3}\,,
$$
thereby normalizing the monopole coefficients to be of the order of unity.
Then, using the single-pole behavior $k^3 \sim k^0 \sim \lambda^{-1}$, we find
$$
b^{00}\sim b^{33} \sim b^{03} \sim 1\,, \qquad
c^{123}\sim c^{012} \sim \lambda \,, \qquad
b^{012} \sim \lambda \,.
$$
The higher multipoles are of the order $\ell^2$, and thus, neglected. This
is a realization of the pole-dipole approximation. In the single-pole
regime, however, only the lowest terms are retained in the limit $\ell\to
0$. This means that, respecting $\lambda\ll\ell$, the terms proportional to
$\lambda$ must also be dropped. As a result, the {\it spin monopole coefficients
$c^{\lambda\mu\nu}$ vanish simultaneously with the orbital dipole
coefficients $b^{\mu\nu\lambda}$}.

The reason for this unusual behaviour is found in the constraint
$c^{012}=2\,b^{012}$. It is obtained by the integration of the more general
relation
\begin{equation}\label{SpinOrbita}
x^{\alpha}\tau^{(0\beta)} - x^{\beta}\tau^{(0\alpha)} =
\sigma^{0\alpha\beta} + div
\end{equation}
that is found to constrain the wave packet currents
(\ref{RnjeTalasnogPaketa}). It relates the wave packet spin to its orbital
angular momentum, so that the expected disappearance of orbital degrees of
freedom in the limit $\ell\to 0$ is followed by the unexpected disappearance
of the spin itself.

To summarize, we see that the spin vector $s^{\mu}$ vanishes in the
single-pole approximation, and the particle trajectory becomes a geodesic
line even in the presence of torsion. The validity of this conclusion,
however, demands some sort of equivalence principle to hold. This is because
the considered wave packet is a solution of the free Dirac equation, and the
inclusion of curvature or torsion may destroy it. What we can do is to
consider weak gravity, so that terms quadratic in curvature and torsion are
neglected. In that case, the free wave packets are a good approximation to
the exact solution, which implies that {\it Dirac point particles behave as
spinless objects in an external gravitational field}. They can still probe
the spacetime curvature, but for the probe of the background torsion, one
needs a thick particle.

\section{\label{ConclusionSection}Concluding remarks}

In this paper, we have considered the motion of point particles with nonzero
spin in spacetimes with curvature and torsion. Using the covariant multipole
formalism developed in Ref. \cite{Vasilic2007}, the world line equations are
derived in the lowest, single-pole approximation. This way, the particle
thickness, and the corresponding internal motion, have been eliminated. Only
mass and spin remained to characterize the particle internal structure.

In our approach, the single-pole behaviour has been adopted for truly
independent variables only. In particular, the antisymmetric part of the
stress-energy tensor has been eliminated from the conservation equations,
prior to imposing the single-pole regime. As a consequence, our single-pole
analysis turned out to differ from the existing literature.

The obtained equations of motion are found to differ from the known
pole-dipole equations by the presence of a novel constraint on the particle
spin tensor. With this constraint, the spin of the Dirac point particle
turned out not to couple to the background curvature, leading to geodesic
trajectories in torsionless spacetimes. In the presence of torsion, however,
a geodesic deviation appears, but also a strong constraint, suggesting that
the spin of the Dirac point particle might be zero. Being a consequence of
our single-pole approximation, this unusual result has been checked by the
explicit construction of the zero-size Dirac particle. To this purpose, a
wave packet solution of the Dirac equation is considered in the limit of
small size and wavelength. The expected single-pole behaviour has been
verified, but also, the spin tensor has been found to disappear in this
limit. In an attempt to explain this unusual behaviour, a relation between
spin and orbital angular momentum has been discovered to hold in our wave
packet example.

Before we close our exposition, let us comment on the possibility that the
disappearance of spin in the zero-size limit might be a general property of
all point particles. First, we notice that there is one more conserved
current in the Dirac theory---the $U(1)$ current $j^{\mu} \equiv
\bar{\psi}\gamma^{\mu}\psi$. In our wave packet example, it is proportional
to the wave vector $k^{\mu}$, and is related to the stress-energy
$\tau^{\mu\nu}$. By a close inspection, we find that the following
manifestly covariant relation holds:
\begin{equation} \label{TauJot}
\tau^{\mu\nu}j_{\nu} \propto j^{\mu},
\end{equation}
Its physical meaning is best seen in the rest frame where it reduces to
$\tau^{\alpha 0}=j^{\alpha}=0$. It tells us that the two currents are
mutually proportional, i.e. that the energy and charge flow in the same
direction. In the limit $\lambda / \ell \to 0$, $\ell\to 0$ it implies the
constraint (\ref{SpinOrbita}), and thus, explains why Dirac point particles
have no spin.

The relations analogous to (\ref{TauJot}) might exist quite generally. There
is nothing special about the statement that all the particle charges flow in
the same direction. This is something one would expect to hold for any type
of point-like matter. However, we need the relation (\ref{TauJot}) to hold
for thick particles, as well. Only then, and only for Dirac matter, the
consequence (\ref{SpinOrbita}) has been derived. If (\ref{TauJot}) were a
general property of the localized matter, the disappearance of spin in the
zero-size limit might be a feature of all massive point particles. Indeed,
when applied to other spins, the relation (\ref{TauJot}) implies
$$
x^{\alpha}\tau^{(0\beta)} - x^{\beta}\tau^{(0\alpha)} =
\sigma^{[\beta0\alpha]} + x^{[\alpha} \del_0 \sigma^{00\beta]} + div \,.
$$
After the integration, the divergence term vanishes, while l.h.s. and the
second term on the r.h.s. give dipole coefficients of the order $\cO_1$. In
the single-pole regime, both disappear, and we end up with the vanishing
spin.

\end{document}